\documentclass{webofc}
\usepackage[varg]{txfonts}   
\begin{document}
%
\title{Licking the plate: dusty star-forming galaxies buried in the ALMA calibration data} 
%
%

\author{\lastname{Jianhang Chen}\inst{1,2}\fnsep\thanks{jhchen@mpe.mpg.de; cjhastro@gmail.com} \and
        \lastname{R.\,J.~Ivison}\inst{1,3} \and
        \lastname{M.~Zwaan}\inst{1} \and
        \lastname{C\'eline P\'eroux}\inst{1,4} \and
        \lastname{A.\,D.~Biggs}\inst{5}
}

\institute{
  European Southern Observatory (ESO), Karl-Schwarzschild-Strasse 2,Garching, Germany 
\and
  Max-Planck-Institut f\"ur Extraterrestrische Physik (MPE), Giessenbachstra{\ss}e 1, Garching, Germany 
\and
  Institute for Astronomy, University of Edinburgh, Royal Observatory, Blackford Hill, Edinburgh UK
\and
  Aix Marseille Universit\'e, CNRS, LAM, UMR 7326, 13388 Marseille, France
\and 
  UK Astronomy Technology Centre, Royal Observatory, Blackford Hill, Edinburgh UK
}

\abstract{%

Deep, unbiased surveys are essential to decipher the cosmic evolution of galaxies.
The submillimetre (submm) and millimetre (mm) windows complement the UV/optical waveband and are key to revealing the cold and dusty Universe.
Traditional ways of conducting deep surveys resort to either lensed fields or target small areas for ultra-long integrations.
These surveys have greatly advanced our understanding of dusty star-forming galaxies (DSFGs), but are susceptible to lensing uncertainties and cosmic variance and will be expensive to expand.
Here, we summarise our recent multi-wavelength survey of DSFGs in the vicinity of ALMA's calibrators: the ALMACAL survey.
These fields have accumulated many hundreds of hours of on-source time, reaching depths and effective areas that are competitive with bespoke cosmological surveys.
We summarise the multi-wavelength number counts from ALMACAL and the resolved fraction of the Cosmic Infrared Background (CIB) from submm to mm wavelengths.
Meanwhile, combining all available ALMA observations in each field results in impressive frequency coverage, which often yields the redshifts of these DSFGs.
The ALMACAL survey has demonstrated the scientific value of calibration scans for all submm/mm and radio telescopes, existing and planned.
}

\maketitle
\section{Introduction}
\label{intro}

The opening of the submillimetre/millimetre (submm/mm) window via various technological developments led quickly to the discovery of dusty star-forming galaxies (DSFGs) \citep{Smail1997, Barger1998, Hughes1998} and, since then, has allowed the further exploration of the early, dusty Universe \
As their name implies, a primary feature of DSFGs is an abundance of dust.
The dust grains, with diameters typically in the range 0.3\,nm - 0.3\,$\mu$m, absorb ultraviolet (UV) radiation and emit mainly in the far-infrared (far-IR) \citep{Galliano2018}.
The far-IR window is not accessible from the ground, which limits studies of DSFGs in the nearby Universe to telescopes in space.
Luckily, the expansion of the Universe shifts the rest-frame far-IR into the submm/mm bands for DSFGs at $z>1$.
In addition, DSFGs with similar star-formation rates (SFR) share similar submm/mm flux densities \cite{Blain93}, even if they are found at a different cosmic epochs. This greatly facilitates the detection of distant DSFGs, pushing out to cosmic dawn in recent years \citep{Cooray2014, Laporte2017, Tamura2019, Inami2022}.
This is critical to our understanding of how the cold gas and star-formation rate density (SFRD) evolve across cosmic time \citep{Tacconi2020, Peroux2020}, as optical surveys are highly biased to UV-bright, unobscured systems.

Deep surveys present one effective way to explore the DSFG population.
Significant steps have been taken by several ALMA Large Programmes.
The first was the ALMA SPECtral line Survey (ASPECS) undertaken in the Ultra Deep Field (UDF) \citep{Walter2016, Gonzalez-Lopez2020}, and the next was towards lensing clusters: the ALMA Lensing Cluster Survey \citep[ALCS --][]{Fujimoto2023}.
These were able to constrain the number counts (number density of DSFGs as a function of their flux density) down to nJy levels.
Meanwhile, they also detected spectral lines and dust continuum from faint high-redshift DSFGs, confronting the formation and evolution of dust in the early Universe \citep[e.g.][]{Laporte2017, Tamura2019}.

Despite extraordinary efforts with all of these ALMA surveys, cosmic variance has remained a fundamental limitation.
For cosmological surveys, there are two major sources of variance.
The first is the Poisson contribution, which dominated the pilot ALMA surveys due to the low number of detections.
The second comes from the clustering of galaxies, where the observed number density will be enhanced if the pointing includes a cluster of DSFGs, or reduced if it includes a void, as is suspected for the UDF.
Gkogkou et al. \cite{Gkogkou2023} quantified the field-to-field variations by incorporating semi-analytical simulations into the largest dark matter simulation (Uchuu). 
Based on their analysis, a minimum survey area of 0.1\,deg$^{2}$ is required to keep the cosmic variance below 20\%, which is significantly larger than all the existing deep ALMA surveys.

Here, we introduce another way to go both deep and wide, exploiting calibration observations.
We first introduce our experience with 10 years of ALMA calibration data.
We present the multi-wavelength number counts of DSFGs and the resolved spectrum of the CIB.
Lastly, we will look ahead to future surveys and how we could we can make the best use of all the data.

\section{Our experience with ALMA calibration observations: ALMACAL}
\label{ALMACAL}

ALMACAL is an ambitious project which exploits the ALMA archive.
Unlike most ALMA projects, which exploit ALMA science scans, with ALMACAL our focus is the calibration scans.
In a typical scheduling block, ALMA observes several bright point sources to calibrate the instrument response as a function of frequency (bandpass), the phase and amplitude variation as a function of time (complex gain), absolute flux density, and sometimes the instrumental polarisation as well.
These calibration scans (sometimes referred to as `overheads') constitute about 20--40\% of the total telescope time.
The quality of these calibration scans influences the quality of the data taken for the primary science target(s).

Calibration observations can also be calibrated for scientific use.
Although each calibrator serves a different role, they can be  calibrated to meet the needs of projects that are independent of their original purpose.
Meanwhile, thanks to the relatively simple spatial structure of the calibrators -- they are selected to be bright and copmpact -- almost all ALMA calibrators are suitable for self-calibration, which corrects any short time variations in their complex gains.

ALMA has accumulated thousands of on-source hours towards its many calibrators, which enables ALMACAL to reach ever deeper.
Meanwhile, by combining all the available data in each field, the improved UV sampling leads to better image fidelity.
In the spectral dimension, the increasing spectral coverage contributes to the spectral-line elements of the ALMACAL survey.
After $\sim$10 years of operations, the depth of ALMACAL in the popular ALMA bands is comparable to the existing ALMA Large Programmes. 
The current status of ALMACAL can be found in \citep{Zwaan2022} and \citep{Chen2023}.

The ALMACAL data workflows, as described in the series of ALMACAL papers \citep{Oteo2016, Klitsch2019, Klitsch2020, Hamanowicz2023, Chen2023}, are highly automated and benefit from the wide applicability of the pipeline.
In summary, each scheduling block was calibrated independently using its own shipped pipeline scripts.
Each field was then self-calibrated twice to improve the phase and amplitude solutions. After each field has been fully calibrated, the calibrator was removed from the field by subtracting a point source model in $UV$ space.
This step is essential, as calibrators tend to variable (they are typically blazars).
After this, the calibrated data are ready for scientific use.

\section{ALMACAL: a multi-wavelength survey for DSFGs}

\begin{figure}[htpb]
  \centering
  \includegraphics[width=0.48\textwidth]{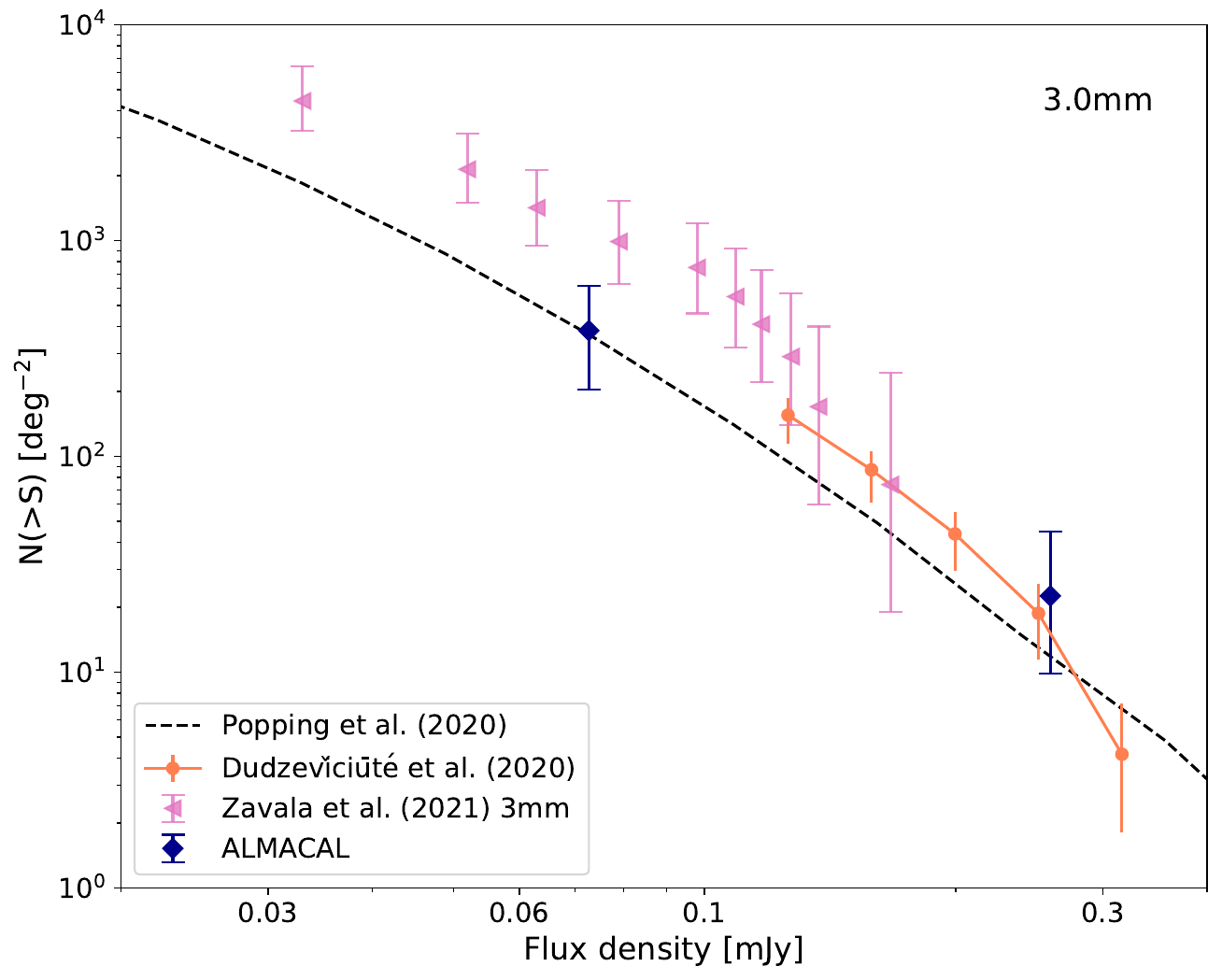}
  \includegraphics[width=0.48\textwidth]{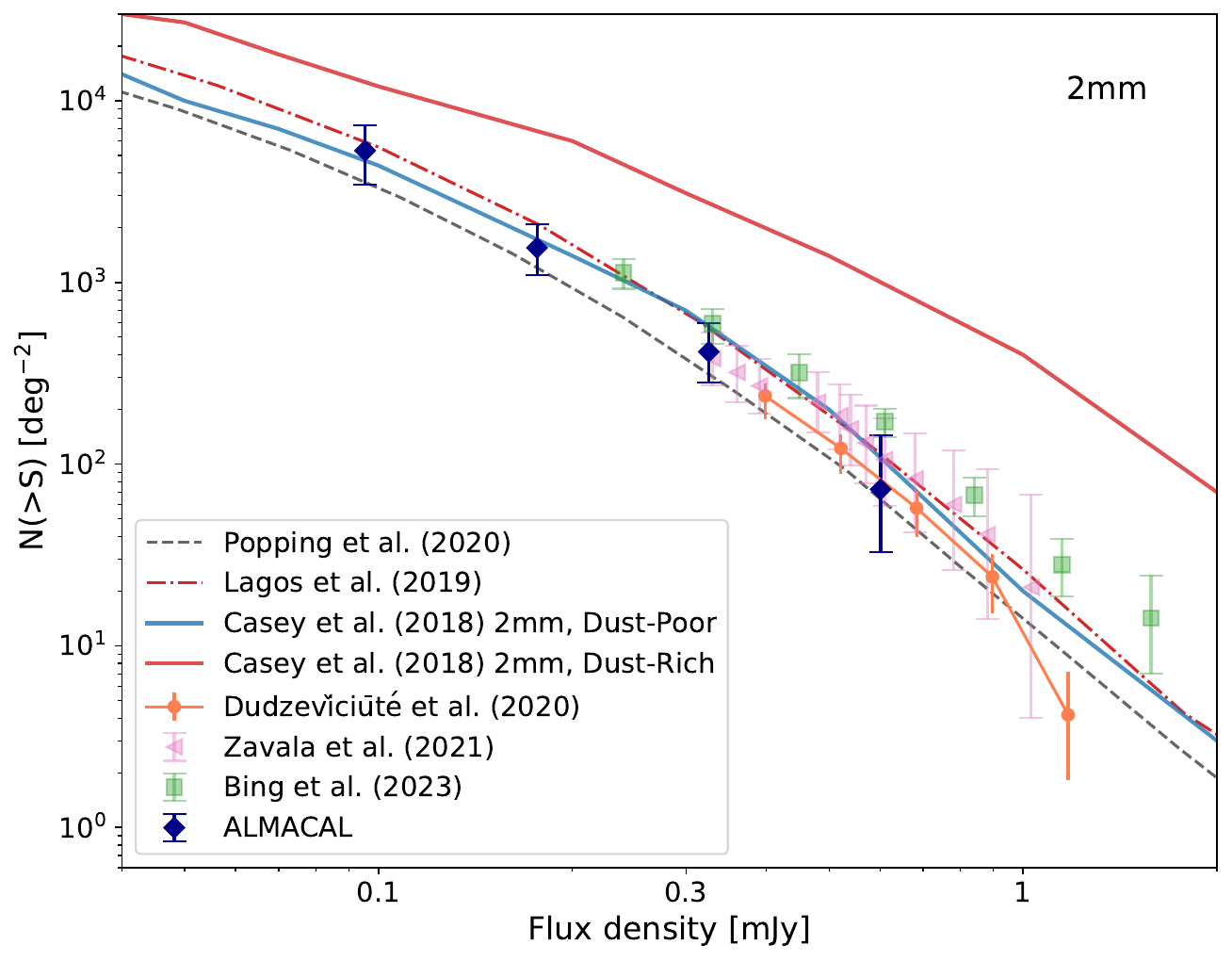}
  \includegraphics[width=0.48\textwidth]{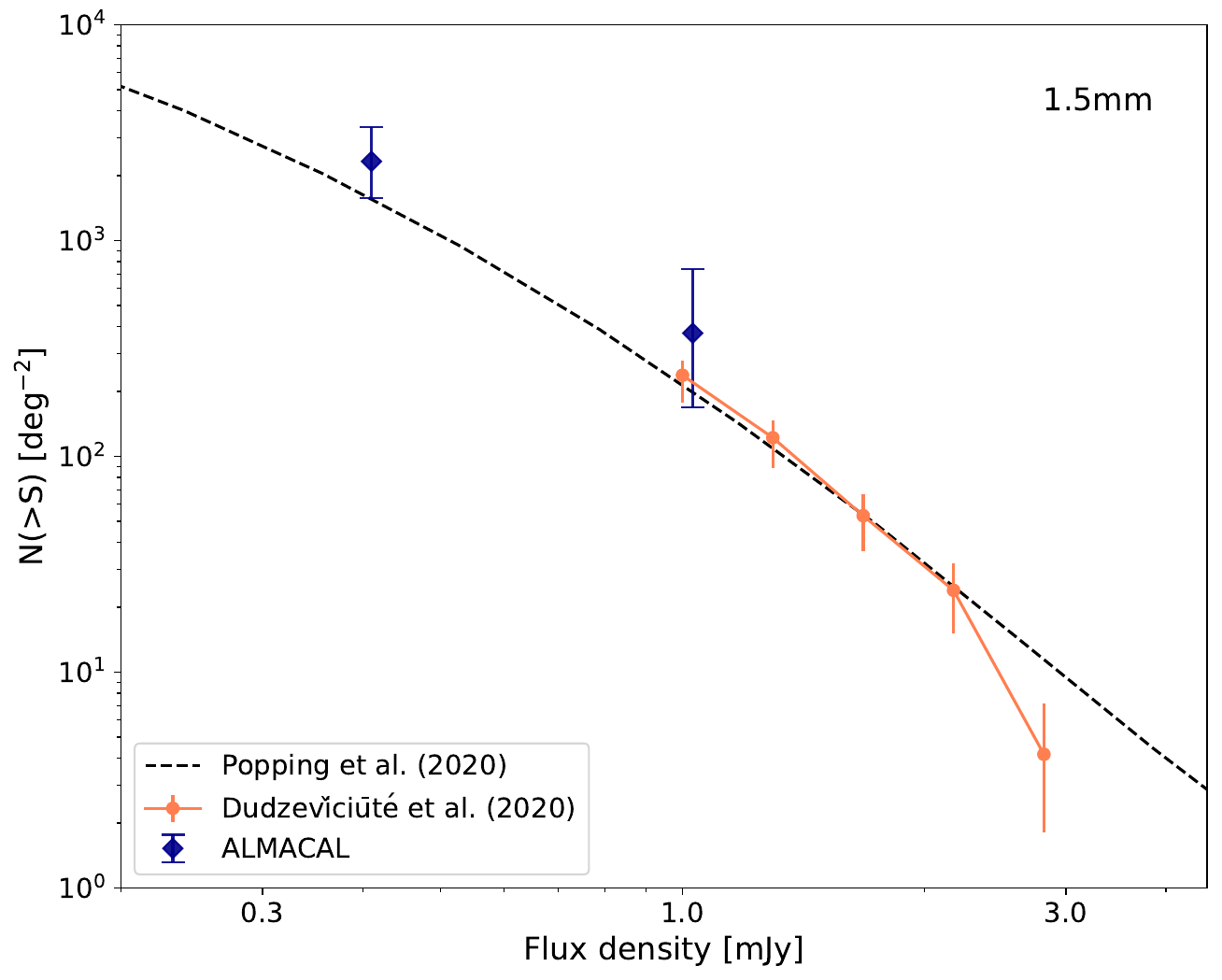}
  \includegraphics[width=0.48\textwidth]{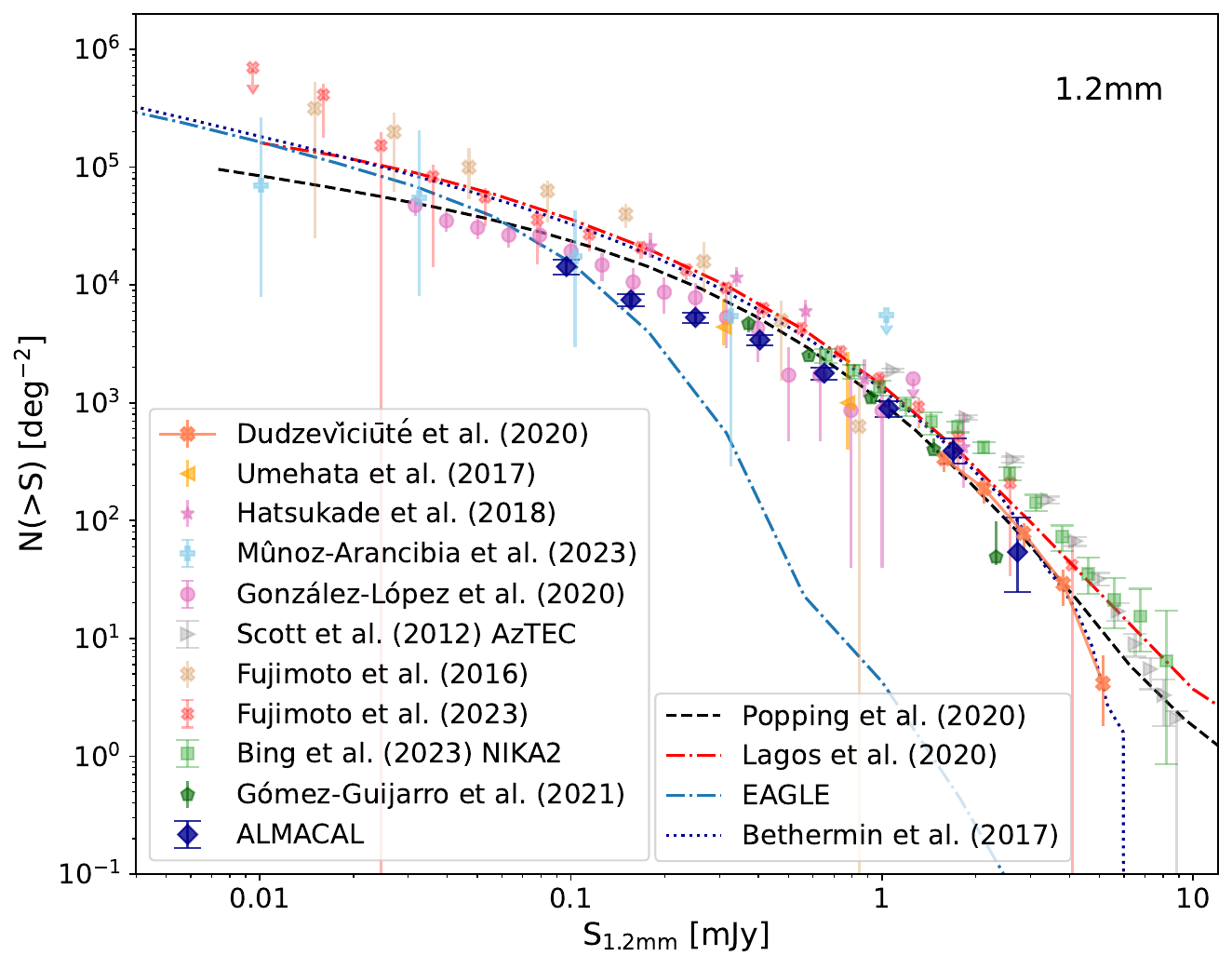}
  \includegraphics[width=0.48\textwidth]{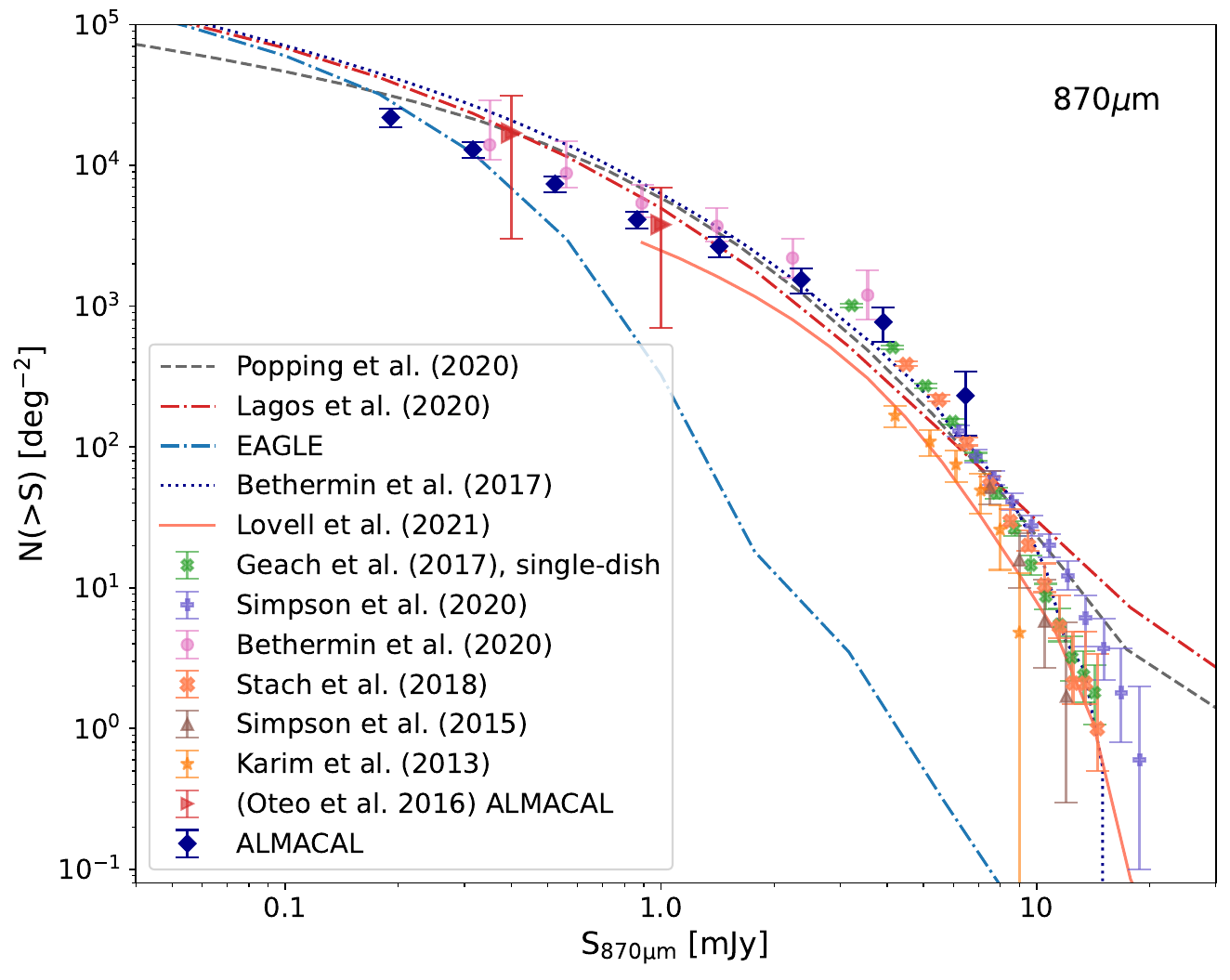}
  \includegraphics[width=0.48\textwidth]{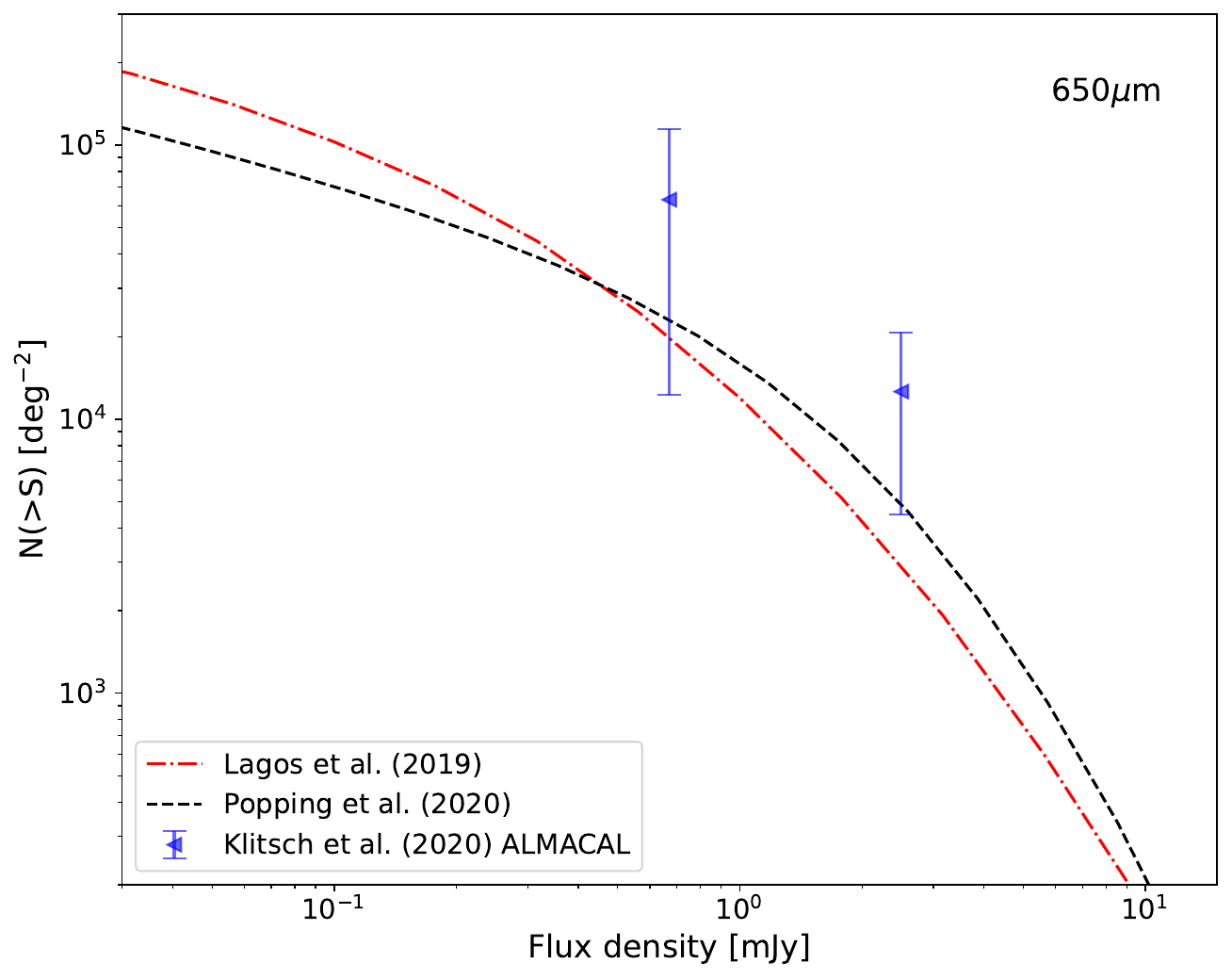}
  \caption{Multi-wavelength number counts of DSFGs in the submm and mm windows. In each subplot, we have included all the recent observational results and model predictions. ALMACAL provides an efficient and economical way of conducting multi-wavelength, deep,  unbiased surveys for DSFGs.}
  \label{fig:almacal_numbercounts}
\end{figure}

\subsection{Multi-wavelength number counts}
Historically, the number counts were mostly determined at wavelengths of 870\,$\mu$m and 1.2\,mm, where the bright dust emission, the reasonable primary beam area and the good atmospheric transmission yield the best compromise to search for DSFGs.
However, the negative $K$-correction varies with wavelength, which makes different ALMA bands sensitive to galaxies at different redshifts and/or with different dust temperatures \citep{Lagache2005}.
For instance, models with different intrinsic luminosity functions at $z>4$ predict considerably different number counts at longer wavelengths \citep{Casey2018}.
This means that a multi-wavelength survey for DSFGs is a great testbed for models of galaxy evolution.

The primary goal of the ALMACAL survey has been to conduct an unbiased, deep survey for DSFGs.
Thanks to the multi-wavelength coverage of the calibrators, it is natural to extend the survey to all the available ALMA bands.
Technical details about the data preparation, cleaning, flux density measurements, completeness corrections, effective wavelength and potential caveats can be found in Chen et al.~\cite{Chen2023} and Klitsch et al.~\cite{Klitsch2020}.
Fig.~\ref{fig:almacal_numbercounts} summarises the available constraints on the submm/mm number counts of DSFGs.
At several wavelengths, such as 1.5\,mm and 650\,$\mu$m, ALMACAL is the first available survey.
At 2\,mm and 870\,$\mu$m, ALMACAL represents the deepest survey available.
These results have demonstrated convincingly that combining calibration scans is an economical and effective way to go both deep and wide, whilst mitigating the problem of cosmic variance.

\subsection{Resolved cosmic infrared background (CIB)}

\begin{figure}[htpb]
  \centering
  \includegraphics[width=0.6\textwidth]{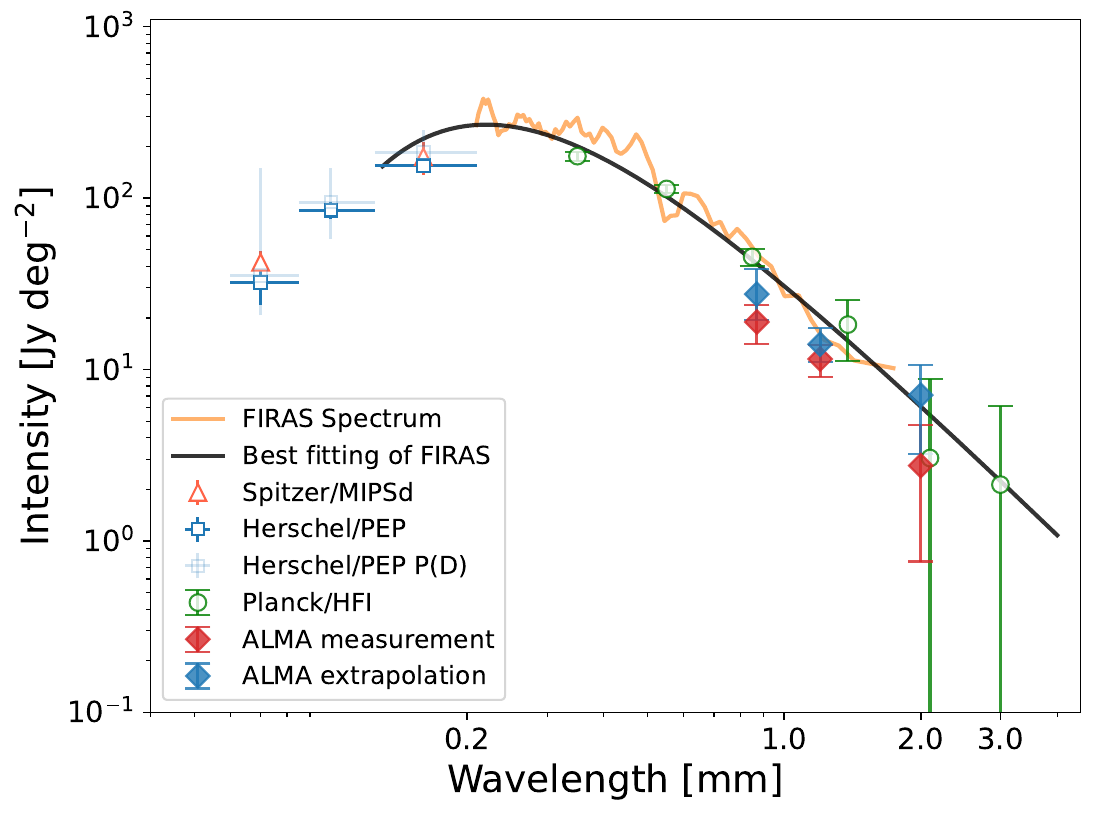}
  \caption{Spectrum of the cosmic infrared background.  The detailed ALMA measurements and extrapolations can be found in Chen et al. \cite{Chen2023} The direct measurement from FIRAS on {\it COBE} is shown in orange. The black curve is the best-fitting FIRAS spectrum from Fixsen et al. \cite{Fixsen1998}. Standalone measurements in the FIR by MIPS on {\it Spitzer} \citep{Dole2006} and using PACS on {\it Herschel} \citep{Berta2011} are also included. We also show the recent results from {\it Planck}/HFI recalibration \cite{Odegard2019}. Compared with the FIRAS direct measurements, ALMA has resolved more than half of the CIB from 870\,$\mu$m to 2\,mm, which suggests that DSFGs are the main contributors to the CIB in the submm/mm range.}
  \label{fig:ALMACAL_CIB}
\end{figure}


The CIB -- covering mid-infrared to
submm/mm wavelengths -- comprises an important component of the energy emitted by galaxies over the entire history of the Universe \citep{Hauser1998, Fixsen1998}.
Along with the optical background, it represents the energy produced by the formation and evolution of galaxies, and all related processes, across all cosmic time.
Since the first measurements of the CIB, a primary goal in astrophysics has been to identify the sources responsible for it, and thus to understand the lifetime energy budget of the Universe \citep{Lagache2005, Dole2006, PlanckCollaboration2014}.

ALMACAL offers a great opportunity to resolve the spectrum of CIB in the submm and mm wavebands.
Combining all the existing surveys reported in Fig.\,\ref{fig:almacal_numbercounts}, ALMA has largely resolved the CIB in three bands (see also Fig.~\ref{fig:ALMACAL_CIB}). 
Based on our best-fit Schechter function, we have extrapolated the total resolved CIB, which is generally consistent with the direct measurement, indicating that DSFGs are the main galaxy population responsible for the CIB in the submm/mm bands.

\subsection{Spectroscopic surveys of DSFGs}

\begin{figure}[htpb]
  \centering
  \includegraphics[width=0.8\textwidth]{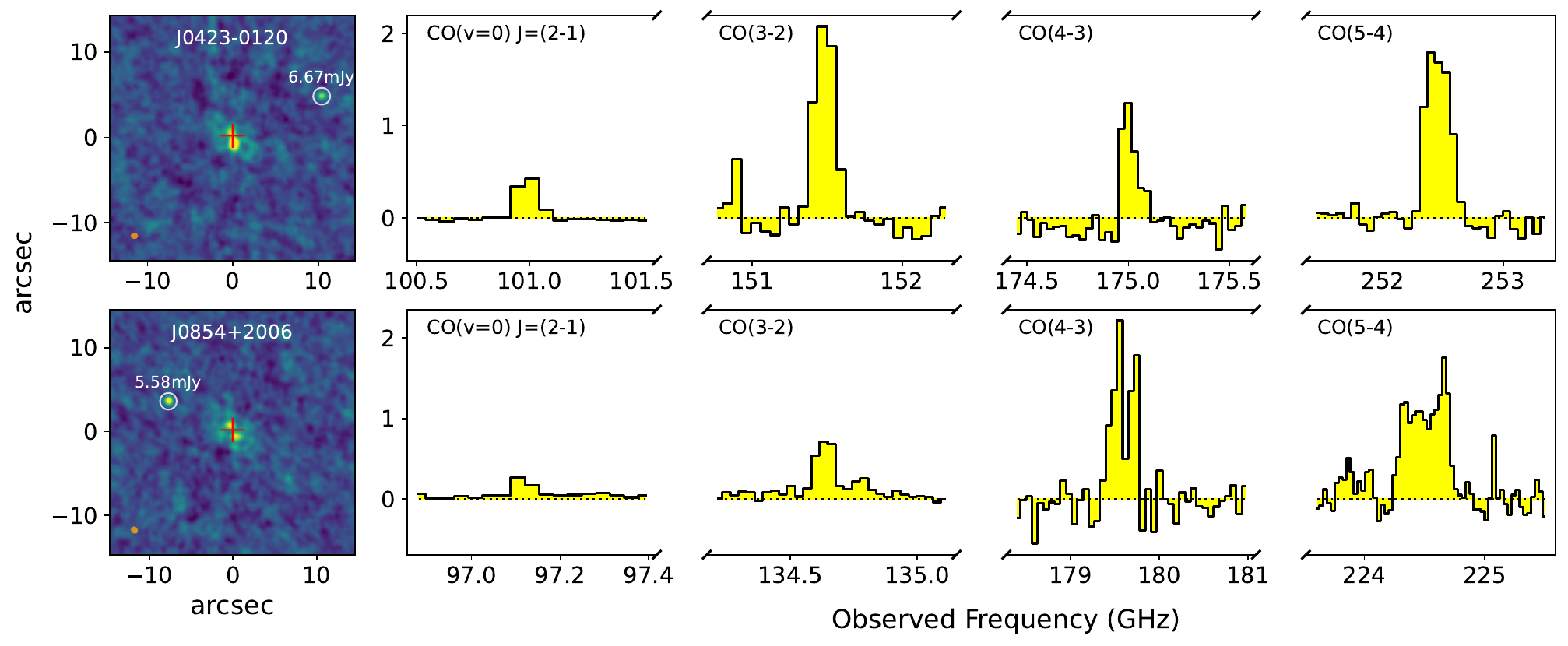}
  \caption{Two bright DSFGs found in ALMACAL have multiple spectral lines confirmed. The first column shows images of the two fields. The white circle marks the DSFG; the red cross is the position of the 
 (removed) blazar calibrator. The second column shows the spectra of the two DSFGs. Both DSFGs are detected in CO $J=2$--1 to $J=5$--4. In J0423$-$0120, the redshift of the calibrator is $z=0.916$, while the redshift of the DSFGs is $z=1.28$. In J0854+2006, the redshift of the calibrator and DSFG are 0.306 and 1.567, respectively.} 
  \label{fig:B7_bright_z}
\end{figure}

Spectroscopic surveys of DSFGs are the next focus of submm/mm cosmology.
Existing surveys with FIR/submm/mm telescopes have discovered thousands of DSFGs.
However, considerably fewer than 10\% of these have known spectroscopic redshifts.
ALMACAL provides complementary way to derive the redshifts of DSFGs. Although a typical ALMA project asks for only one spectral window, or perhaps a small handful,
the diverse range of ALMA projects overall mean that there is demand for many different spectral windows, which leads to extremely wide spectral coverage of the calibrators, which is very handy when searching for spectral lines from the DSFGs in those fields.
In Fig.~\ref{fig:B7_bright_z}, we show two of the many DSFGs discovered in the ALMACAL survey.
Both have serendipitous detections of multiple CO transitions. 
Another example was reported in Chen et al. \cite{Chen2023a}, which confirmed the redshifts of five DSFGs in a single field around the blazar, J0217$-$0820.
With the future ALMA upgrade to a considerably wider instantaneous bandwidth, ALMACAL will become even more efficient in searching for faint DSFGs and determining their spectroscopic redshifts.


\section{Summary and future prospects}

Calibration observations are essential to all telescopes.
These observations often contain much more information than was required to meet their original purpose.
Particularly at submm/mm wavelength, these calibration data have proved suitable to conduct a deep and wide survey for DSFGs.

ALMACAL has taken advantage of all the archival ALMA calibration observations with which we have conducted a multi-wavelength survey for DSFGs and resolved a significant fraction of the CIB at submm/mm wavelengths. We have also demonstrated the usefulness of the very wide spectral coverage that is typically available for each ALMA calibrator.
While ALMACAL is related to the ALMA calibration data, a similar approach can be applied to other interferometers and even single-dish telescopes.

Calibration observations have also proved suitable for many other scientific projects, including the study of blazar variablity, of course.
Since these calibrators are exceptionally bright, they have also proved ideal places to search for absorbers in their line of sight \cite{Klitsch2019}. 
We recommend that a systematic approach allowing the best use of calibration observations should be considered for all future telescopes or upgrades.

\end{document}